# Benchmarking the Privacy-Preserving People Search

Shuguang Han, Daqing He and Zhen Yue
School of Information Sciences, University of Pittsburgh
135 N Bellefield Ave., Pittsburgh, PA, United States
shh69@pitt.edu, dah44@pitt.edu, zhy18@pitt.edu

## ABSTRACT
People search is an important topic in information retrieval. Many previous studies on this topic employed social networks to boost search performance by incorporating either local network features (e.g. the common connections between the querying user and candidates in social networks), or global network features (e.g. the PageRank), or both. However, the available social network information can be restricted because of the privacy settings of involved users, which in turn would affect the performance of people search. Therefore, in this paper, we focus on the privacy issues in people search. We propose simulating different privacy settings with a public social network due to the unavailability of privacy-concerned networks. Our study examines the influences of privacy concerns on the local and global network features, and their impacts on the performance of people search. Our results show that: 1) the privacy concerns of different people in the networks have different influences. People with higher association (i.e. higher degree in a network) have much greater impacts on the performance of people search; 2) local network features are more sensitive to the privacy concerns, especially when such concerns come from high association peoples in the network who are also related to the querying user. As the first study on this topic, we hope to generate further discussions on these issues.

## Categories and Subject Descriptors
H.2.8 [Database Applications]: Data Mining; H.3.3 [**Information Storage and Retrieval**]: Information Search and Retrieval - Search process

## Keywords
People Search; Privacy-preserving networks; Privacy-preserving people search

## 1. INTRODUCTION
Modern search engines often assume that their search algorithms should return the most relevant documents to a query. However, in many occasions, users actually want to look for relevant people rather than documents. For example, company recruiters may need to find appropriate job candidates for a job opening [1]; or conference chairs may need to invite the right experts to form a program committee [2]. These topics have been studied as the expert finding problems in the information retrieval community [3], and the expert is often defined as the people who have domain knowledge for a given topic. However, expert finding is only one type of people search tasks. Many other scenarios such as finding appropriate collaborators [4] or thesis committee members [5], require not only the topical expertise matching but also the social matching [6] because a higher social similarity make it easier for people to connect.

In order to perform social matching, the retrieval systems need to access users' social networks and return the potential candidates who have either direct or indirect connections with the given users. However, privacy has been identified as a major concern in many social network services [7, 8] - users often either opt out from certain social networks or provide incomplete or even fake social network information. Early research work has shown that many data mining algorithms may not work or even harm user experience when equipped with such incomplete and noisy social information [9]. Recently, researchers start to incorporating social information into people search systems, and the coauthor networks generated from scholarly publications were often utilized [4, 5, 10]. However, probably because the coauthor networks often have less privacy concerns, little attention has been paid to the privacy related issues in people search. Furthermore, there is no study on how incomplete social networks would affect the performance of people search systems.

In this paper, we are particularly interested in the privacy issues in people search and the impacts of these issues on people search performance. The TREC experience demonstrates that it would be a critical drawback for studying the search problems if there are no appropriate test beds. Considering the difficulty of obtaining an open privacy-concerned social network and the expense of constructing such a network from scratch for research purpose, we propose in this paper to simulate the privacy-concerned social network using the public available coauthor networks. Note that, users in many social network services are able to keep both their profiles and social connections as private. In this paper, we focus on the privacy issues of sharing social connections.

The key assumption of our simulation is that a coauthor network would have the same or similar network characteristics with a privacy-concerned social network. The foundation of our simulation approach is based on some existing studies [11-13], which state that many real-world social networks (including coauthor networks and many other privacy-concerned networks such as Facebook social networks) share the same patterns: they are small-world networks and their degree distributions are highly skewed. Newman [14] studied the assortative patterns (the preferences of connecting people who share the similar features) of social networks. He found that the social networks showed assortatively mixed patterns, whereas technological and biological seems to be disassortative. Therefore, it is reasonable to assume that coauthor networks and many privacy-preserving networks (because they are both social networks) share some important common characteristics. Therefore, coauthor networks, which are publically available, can be used as the surrogate for studying privacy-preserving social networks. In the remaining part of this paper, all the privacy related discussions are based on coauthor network and coauthor network-based people search.

In order to study the impact of privacy concerns to the people search performance, we need to examine how the social network information is used in existing people search systems. We refer the global network features as the features that are propagated through the whole networks while the local network features are those that are directly related to the ego-network of the querying user [15]. Some people search systems adopted only the local network features [4], whereas some others used both the local and global network features [5, 10]. For example, Han et al. [5] took into consideration of both the local social similarity between the querying user and each returned candidate (measured by the proportion of common social connections) and the global authority of each returned candidate (measured by the PageRank value running on the whole social networks). They found that combing both global and local network features with the topic relevance would provide better support of modeling diverse people search contexts and further augment the search experiences. Since both the global and local network features played important roles in people search systems [5], the study of privacy needs to consider both.

Both the local and global network features could be influenced by the completeness of social network information. Therefore, a privacy-preserving network with many private (unrevealed) social connections would affect the calculation of the global and local network features, which may in turn affect the people search performance. The incomplete social contexts of the querying user and the network candidates affect the calculation of the proportion of common coauthors between them. This is the reason why we examine the impacts of privacy concerns on the local network features for both network candidates and querying user. When analyze the local network features, we study the privacy settings of querying users and candidates separately. The global network features rely on the information propagation through the whole network which is only related to network candidates. We study global network features for network candidates only.

In summary, we identify that privacy-preserving people search is still an almost untouched research topic. In this paper, we make the first attempt to provide some benchmarks by simulating privacy-preserving networks and examining how these networks affect the performance of people search. To achieve the goal of this study, we need to properly simulate different types of privacy-preserving networks. A privacy-reserving network is essentially a subset of the full network, so we model different privacy concerns as different sampling strategies (the purpose is to sample a subset of privacy-concerned people). We discuss sampling strategies in section 2. To be specific, our research questions are:

- RQ1: How to properly simulate different types of privacy-preserving social networks?
- RQ2: How does each type of privacy-preserving network affect the global and local network features?
- RQ3: How does the global and local network features derived from privacy-preserving networks further affect the people search performance?

## 2. DATASET AND METHODOLOGY

### 2.1 Experiment Dataset

Our experiments in this paper reuse the user study data and the publication collection presented in Han et al. [5]. The dataset used in that study was an academic publication collection containing 219,677 conference papers from the ACM Digital Library. These papers were published in academic conferences (the full list of conferences is available at ACM Digital Library [1]) between 1990 and 2013. Only public available information of a paper (the title, abstract and authors) was collected. The unique identifier assigned by ACM Digital Library was used to identify each author, and no further author disambiguation step was performed. In total, the collection contains 253,390 unique authors and 953,685 coauthor connection instances. Therefore, that collection contains both content information about papers (title and abstract) and social network of authors (i.e., coauthor networks).

The goal of the user study presented in Han et al. [5] was to evaluate a people search system. The study involved four different people search tasks, each of which aimed to search for 5 candidates satisfying a querying user's search need. Two systems were used in the study: a baseline plain content-based people search system and an experimental system that enhances people search with three interactive facets: content relevance, social similarity between the user and a candidate (the local network feature) and the authority of a candidate (the global network feature). The experiment system allowed the querying users to tune the importance associated with each facet in order to generate a better candidate search result. 24 participants were recruited for the user study. At the beginning of the user study, each participant was asked to provide their publications and their social connections (such as advisors). In the post-task questionnaire, the participants were asked to rate the relevance of each marked candidate in a Five-point Likert scale (1 as non-relevant and 5 as the highly relevant).

We reuse the data from [5] in the following ways. First, we use the same academic publication collection which contains both the papers and the coauthor networks. Secondly, we use the marked highly relevant candidates (i.e., those with ratings higher than 3) from the user study as our ground-truth, which are further used to measure the effectiveness of the search algorithms under different privacy-preserving network scenarios.

### 2.2 Configuring Privacy-Preserving Networks

We identify two different types of users in our study, the people who initiates the people search requests (i.e. the participants in the user study. Therefore, they are called querying users) and the candidates in the publication collection and the coauthor networks (therefore, called the candidates). We treat them differently because: 1) although many querying users would be on the coauthor network, some others may not be; 2) more importantly, we believe that the calculation of local network features can be influenced by the privacy settings of the querying users as well as the candidates, and the impacts of privacy setting from different users would be different.

#### 2.2.1 Modeling Privacy for the Candidates

Although the privacy settings are related to various factors, those factors would result in a common outcome – a user either has privacy concern or not. We assume that there is a probability (i.e. $p_i$) for each candidate being privacy-concerned. Based on different roles that people can play in a network, we think that modeling privacy concern as being associated with the candidate's degree of associations (i.e. the coauthor relationships) on the network would be a reasonable approach to study the impacts of privacy settings for people with different roles. We could see that there are two extremes for different candidates to have privacy concerns: 1) the top degree of association candidates have privacy-concerns; or 2) the bottom degree of association candidates have privacy concern.

Suppose that for each candidate $i$, his/her degree of association on the network is $d_i$ and the maximized degree on the network is $d_{max}$, we have Eq. 1 to provide one formula with a parameter $\lambda$ for modeling candidates with different degree of associations on the network to have privacy concerns. When $\lambda$ is set as negatives or positives, we can obtain different simulations for indicating either top-degree or bottom-degree candidates to have more privacy concerns. The absolute value of λ corresponds to the power of emphasizing on top-degree or bottom-degree candidates. When λ is set to 0, it is uniform and each user has equivalent probability.

---

[1] http://dl.acm.org/proceedings.cfm

$$p_i \propto \left(\frac{d_i}{d_{max}}\right)^\lambda \qquad \text{Eq. 1}$$

Besides λ, we need another parameter to control the proportion of candidates on the networks who have the privacy concerns (noted as $p_b$). In this paper, we will test nine different $p_b$ (from 0.1 to 0.9, with 0.1 for each step) and under each $p_b$. Besides, we also test different values of λ. For each pair of <λ, $p_b$>, we sample 10 different runs to remove the bias. Our reported results are based on the average over those 10 runs. To be specific, suppose that we have N candidates and we think that N × $p_b$ of them have privacy concern. The goal of sampling, therefore, is to return N × $p_b$ sampled privacy-concerned candidates. Our sampling algorithm is a "*sampling without replacement*" (see Figure 1).

---

Algorithm: Sampling privacy-concerned candidates

---

**Input**: N, $p_b$ and λ; **Output**: N × $p_b$ privacy-concerned candidates U

---

Procedure:

1 : compute $p_i$ using Eq. 1, put it in array P[] and compute the sum S of P[]
2 : for run = 1 : 10
3:     M = N
3 :    for i = 1 : N × $p_b$   //sampling N × $p_b$ candidates
4 :        randomly generate a number r in [0,S)5:
5 :        for a = 1 : M
6 :            if Σ P[a] ≥ r
7 :                put the corresponding candidate into U
8 :                S = S – P[a];
9 :                break;
10:    M = M -1;

---

Figure 1: Algorithm for generating the privacy-concerned candidates

### 2.2.2 Modeling Privacy for the Querying Users

The local network feature in this paper refers to the proportion of common social connections between the querying users and the existing candidates. Therefore, the privacy settings of both people will influence the calculation of the local network feature. Modeling privacy concerns for the candidates has been discussed above; here we present our modeling of the privacy concerns on the querying users. The social connections of the querying users were obtained through the users themselves in the user study (more details see Han et al. [5]). In that study, each participant was asked to provide his/her personal information as well as his/her close social connections.

The privacy-conscious users may either do not provide any or only provide incomplete personal social information. In our study, therefore, we introduce the completeness of the provided information ($p_c$) as the indicator of the querying user's privacy concerns. It is measured by the percentage of social connections that a querying user provided over the complete "oracle" social connections of that user. The "oracle" social connections are simulated by the user provided information from the user study in [5] because the users were explicitly asked to provide complete social connections during the user experiment.

In this paper, we test elven different values of $p_c$ (from 0.0 to 1.0, with 0.1 for each step). Note that, when set $p_c$ = 0.0, it corresponds to the scenarios that we do not have any social information for the querying user. When we set $p_c$ = 1.0, it means that the complete social connections for the querying user is available. When set $p_c$ to the other values, we can only use partial social connections. To remove the sampling bias, we randomly sample the incomplete social connections 10 runs and the reported results are based on the average over 10 runs.

### 2.3 Experiment Setup

Our study involves two sets of experiments. The first set examines the impacts of various privacy settings on the computing of global and local network feature. The second set tests their further influences on people search.

### 2.3.1 Testing the Impacts on Global Network Feature

Since the local network feature is directly related to the querying users, it is difficult to study it independently. In contrast, the global network feature is computed through the propagation on the whole network and it is independent of the querying users. So, we only examine the influences of different privacy settings on the global network feature in this section.

The global network feature of a candidate is represented as his/her authority value, which is measured by the PageRank value on the coauthor networks. We first compute the authority value ($pr_a$) for each candidate *a* using the whole network information. This is treated as the ground-truth values. To test the impact of a privacy setting, we re-compute the authority value ($pr_a^p$) for the candidate *a* with different portion of people on the network do not share their social connections because of the privacy concerns. We use the Mean Absolute Error (MAE) between the new authority values and ground-truth authority values over all of the authors as the indication of the impact from privacy concerns (see Eq. 2).

$$MAE = \frac{1}{N}\sum_{a=1}^{N}|pr_a - pr_a^p| \qquad \text{Eq. 2}$$

### 2.3.2 Testing the Impacts on People Search

When examining the impacts of different privacy-preserving networks on the people search performance, we adopted the user study data from Han et al. [5]. In that experiment setting, the effectiveness of a people search was affected by three facets: content relevance, local network feature and global network feature. The three facets are displayed to the querying users so that the users could directly configure the importance of each facet. To test the influences of using privacy-preserving networks, we can directly test its impacts on a live system by comparing system performance in two scenarios: one with complete network and the other one with privacy-preserving networks. However, it will be very time-consuming and may be unable to detect the subtle differences. Therefore, we decide to conduct a simulation study based on the queries and marked candidates from [5].

We assume that a querying user *u* issued several queries in order to finish a task and under *K* queries, *u* has marked at least one candidate. We name those K queries as the *effective queries*. We assume that the purpose of each effective query is to retrieve the best-matched candidates (i.e. the ground-truth). Although the ordering of those *K* queries may reveal their importance in the whole search process, we do not consider such information in this paper for simplicity. Therefore, for each effective query in [5], we compute three scores: the query-candidate content match $S_C$, the local network feature $S_L$ and the global network feature $S_G$. Those scores were transformed into logarithmic values and combined linearly. In a live system, the querying users can tune the importance of each facet: $w_c$ (for $S_C$), $w_g$ (for $S_G$) and $w_l$ (for $S_L$).

The computation of each score and their integration are the same as Han et al. [5]. The Integration score S is computed using Eq. 3. The candidates are ranked based on this score.

$$S = w_c S_C + w_g S_G + w_l S_L \quad \text{Eq. 3}$$

For each effective query, different configurations of $w_c$, $w_g$ and $w_l$ yield different search performance. Lacking of the real user interactions, we cannot obtain how users would set those weights. In the simulation study, we assume that users are able to tune the best configurations to achieve the best search performance. The search performance of each effective query $q_i$ is measured by the Average Precision (AP) under the best configuration of $w_c$, $w_g$ and $w_l$, as shown in Eq. 4. The AP is computed using the ground-truth data (the marked candidates for a task with ratings bigger than 3.0) from the user study in Han et al. [5]. The ground-truth is built for user-task pair so that any of the $K$ effective queries within one user-task pair would share the same ground-truth. The search performance of each user-task pair is then measured by the Mean Average Precision (MAP) over all of the $K$ effective queries, as shown in Eq. 5. Then, the search performance depends on the $S_C$, $S_G$ and $S_L$ (as shown in Eq. 3.), which are determined by the available information in the privacy-preserving networks. The comparison of privacy-preserving networks can be transformed to compare the MAP.

$$AP_{q_i} = \max_{w_e, w_p, w_s \in [0,1]} AP_{q_i}(w_e, w_p, w_s) \quad \text{Eq. 4}$$

$$MAP = \frac{1}{K} \sum_{k=1 \dots K} AP_{q_i} \quad \text{Eq. 5}$$

## 3. IMPACTS ON GLOBAL FEATURE IN PRIVACY-PRESERVING NETWORKS

In this section, we study how different privacy-preserving networks influence the computation of the global network feature and how it further affects the performance of people search.

### 3.1 Impacts on Global Network Feature

We simulate different privacy-preserving networks by setting different $\lambda$ in Eq. 1. We compare five $\lambda$ in this paper: -1.0, -0.5, 0.0, 0.5 and 1.0. Under each $\lambda$, we then adopt the sampling procedure described in the section 2.2.1 to choose a certain percentage ($p_b$ in the Figure 1) of privacy-concerned candidates. To measure its impacts on the computing of global network feature, we measure the MAE between its values on the full networks and the sampled privacy-preserving networks.

The MAE results are shown in Figure 3. As stated, when $\lambda$ is set to 0.0, the candidates on the network have uniformed probability ($p_i$ in Eq. 1) of being concerned on sharing social connections. We treat it as one of the baselines. We also set $\lambda$ as negative values to simulate the scenario that candidates with low association degrees have more privacy-concern. Since those low association degree candidates only affect a small proportion of the connections on the network, we suspect that they have less impact. The results from Figure 2 confirm our expectation. In addition, $\lambda$ with smaller negative values (i.e., bigger absolute values) results in slightly better MAE, which is not surprising based on our suspect.

When set $\lambda$ into a positive value, it corresponds to the scenario that the high association degree candidates have higher chance to have privacy concerns. Since those high association degree candidates are usually well-connected in networks, we anticipate a higher impact from their privacy concerns. We see in Figure 2 that the MAE curves for both two positive $\lambda$ values are above the baseline. When sampled more privacy-concerned candidates from high-degree candidates (i.e. compare $\lambda = + 0.5$ with $\lambda = + 1.0$), we see an increase of the MAE errors.

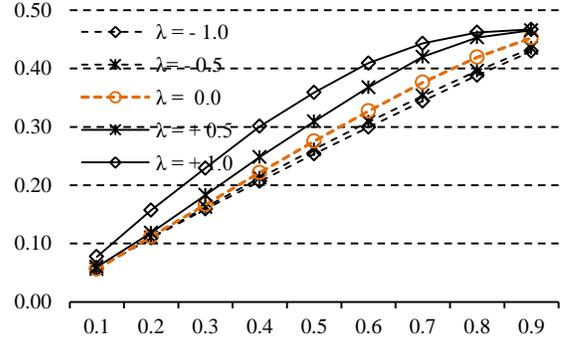

Figure 2: The impacts of different privacy-preserving networks on the calculation of global network feature. We measure the impacts using MAE. **X axis**: $p_b$ in Figure 1; **Y axis**: the MAE. Each value is aggregated over 10 runs. (MAE, the smaller the better)

### 3.2 Impacts on the People Search

We further study how different privacy-preserving networks affect the people search performance. We took $\lambda = -1.0$ (+1.0) as the upper (lower) bound based on the result in Figure 2 and still used $\lambda = 0.0$ as the baseline.

To simulate and measure the people search performance, we need to set appropriate parameters ($w_c$, $w_g$ and $w_l$) in Eq. 3. Since we only focus on the impact of the global network feature in this section, we set the weight for local network feature $w_l = 0$. We estimate the parameters based on the full network information, and assume that parameters are also applied to privacy-preserving networks. We acknowledge the limitation of not tuning parameters for each network. We think the parameters reveal users' objective view of the importance of each facet and it remains the same under different networks. The parameters we used in this section are $w_c = 1.0$ and $w_g = 0.1$.

The MAP evaluations under different privacy-preserving networks (different values of $\lambda$ and $p_b$) are shown in Figure 3. We also plot the MAP performance using the full network information (the red solid line) as an upper bound baseline. We find that the results of $\lambda = -1.0$ have very similar performance to the upper bound baseline even when $p_b$ is as large as 0.9. This is because here only those low-degree candidates have privacy concerns while the core candidates with medium or high degree remains in the network. In contrast, the results of $\lambda = +1.0$ (high-degree people has more privacy concerns) have clearly impacts on the people search performance even when $p_b$ is as small as 0.1 and 0.2. This is because many core candidates with top degree of associations are removed from the networks.

Although the maximal change of MAP is a 3.87% drop (relative percentage when $\lambda=+1.0$ and $p_b=0.8$, comparing to the "Full Networks"), the changes for all $p_b$ are still significant under the Wilcoxon Sign Test (e.g. p-value=0.040 for $p_b=0.1$, p-value =0.016 for $p_b = 0.2$ and p-value= 0.000 for $p_b=0.3$ and etc). Again, the results of $\lambda = 0.0$ lie between that of $\lambda = + 1.0$ and that of $\lambda = -1.0$ because the high- or low-degree candidates have the same probability of being sampled as the privacy-concerned candidates.

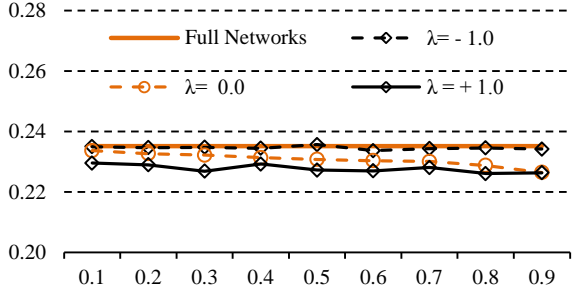

Figure 3: The impacts of new global network feature under different privacy-preserving networks to the performance of people search. The impact is measured by MAP. **X axis**: $p_b$ in Figure 1; **Y axis**: the MAP. Each value for different λ (except the "Full Networks") is aggregated over 10 runs. (MAP, the bigger the better)

## 4. IMPACTS ON LOCAL FEATURE IN PRIVACY-PRESERVING NETWORKS

In this section, we try to understand the impacts of privacy-preserving networks on local network feature. Since it is related to both the candidates and the querying users, we study the privacy settings for both two types of users.

### 4.1 Impact of Candidates' Privacy Setting on Local Network Feature

Since we are focusing on the local network feature in this section, we set $w_g = 0.0$. To find appropriate weights for $S_C$ and $S_L$ (i.e. the optimal $w_c$ and $w_l$), we re-examine users' people search process based on the user study data and find the corresponding optimal parameters that maximize the people search performance over all effective queries. Same as the Section 3.2, in this process, we use the full network information and assume that the same parameter setting also applies in the privacy-preserving networks. The best parameters we chose is the $w_c = 1.0$ and $w_g = 0.082$. We also use the same parameters in the section 4.2.

The MAP evaluations on different privacy-preserving networks are shown in Figure 4, where we examine the results of three different λ values: -1.0, 0.0 and +1.0. Besides, we consider the "Full Networks" as an upper bound baseline. It is the same as what we did in Section 3.2. We find that local network feature produces more improvements on the performance of people search than global network feature -- the MAP equals to 0.2352 for global network feature (combing with the content relevance) while it equals to 0.2752 for local network feature (combing with content relevance) when using the full network information. The difference is significant under the Wilcoxon Sign test, p=0.003. However, we observe that local network feature is more sensitive to the privacy setting than global network feature – the maximized MAP change for the λ = 0.0 is less than 0.01 for global network feature (as shown in Figure 3) while it changes more than 0.035 for local network feature (as shown in Figure 4).

We further find that removing those high-degree candidates (i.e., λ=+1.0) has a great impact -- the performance has a substantial drop even when only a small portion of candidates have privacy concerns ($p_b$ =0.1 or 0.2). This indicates the import roles that the high-degree candidates played in the computing of local network feature. We think it may be because of that most of the desired candidates (i.e. candidates in the ground-truth) for our user study are actually directly or indirectly connected to the top degree candidates. However, this is not the case when λ=-1.0 where less well-connected candidates are removed. The MAP of randomly selecting candidates (λ=0.0) to have privacy concerns lies between that of λ=-1.0 and that of λ =+1.0.

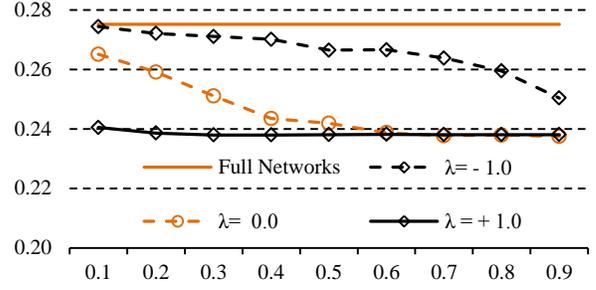

Figure 4: The impacts of new local network feature under different privacy-preserving networks to the performance of people search. The impact is measured by MAP. **X axis**: $p_b$ in Figure 1; **Y axis**: the MAP. Each value for different λ (except the "Full Networks") is aggregated over 10 runs. (MAP, the bigger the better)

### 4.2 Impacts of Querying Users' Privacy Setting on Local Network Feature

The last privacy setting we examined is related to the completeness of social information provided by the querying users that is to test the influence of different settings of $p_c$ (see the section 2.2.2 for its definition) on people search performance. The MAP evaluations over different $p_c$ are shown in Figure 5. The "No Social Info." means that we do not use the local network feature. The "Full Social Info." corresponds to the scenario that we can obtain the complete user social connections and use them to compute the local network feature. The "No Social Info." performs as the lower bound of the MAP whereas the "Full Social Info." acts as the upper bound.

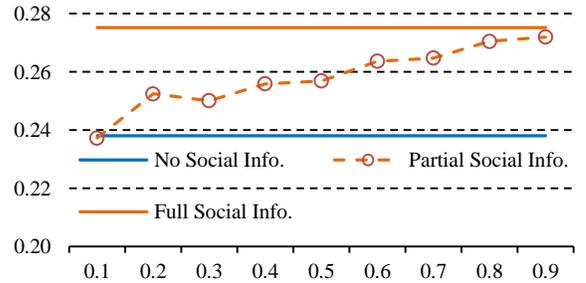

Figure 5: The impacts of new local network feature under different privacy settings of querying users to the people search performance. **X axis**: $p_c$, i.e. the completeness of user provided social information; **Y axis**: the MAP. Each point for the "Partial Social Info." is averaged over 10 runs. (MAP, the bigger the better)

We observe that the upper bound is significantly better than the lower bound (+15.58%, with p-value= 0.001 under Wilcoxon Sign Test), which indicates the usefulness of involving local network feature of the querying users in the people search process. We also find that the search performance will keep steadily increasing when having more social information about the query user (the dotted red line with "Partial Social Info.").

## 5. CONCLUSION AND DISCUSSIONS

People search has been extensively studied in recent years. Many of the researchers identified that social network information is an important resource for improving the people search performance

[4, 5, 10, 15]. The social networks can be used to infer the local network feature between the querying users and candidates, as well as the global network feature regarding the candidates. However, both the local and global network features can be highly affected by the privacy settings of querying users and candidates. Although the privacy issues are increasingly important in recent years, its impacts on people search haven't been studied yet.

It may be due to the difficulty of obtaining a privacy-preserving social network and make it openly available for research purpose. Therefore, in this paper, we focus on simulating the privacy-preserving social networks using a publicly available network – the academic coauthor network. The privacy could come from either the querying users or the candidates in the networks. We studied their impacts separately. For the querying users, we treated the completeness of social information as a parameter to simulate the scenario that users do not provide full social information. For the candidates, we introduced the proportion of candidates that has privacy concerns and the strength of association (i.e. his/her degree in the networks) as two parameters. We assume that candidates' privacy concerns are correlated with their degree of association in networks.

When using the full network information, we find that both the local and global network features provide significant boosts on the performance of people search (compare to not using social network). However, comparing to the global network feature, the local network feature can provide greater improvements. Using the simulated networks, we also find that privacy-preserving networks have significant influences on the performance of people search with both the local and global network features (comparing to the use of complete network information).

In additional, we observe that different roles of candidates can exert different impacts on the computing of global network feature and they further impose different influences on the people search process. The privacy concerns from the high-degree candidates in the network have more impacts. Since the local network feature is related to both the querying users and the candidates in the networks, we find that the privacy concerns from both of them have significant impacts on the search performance. The privacy concerns from high-degree candidates have bigger influences on the people search than that of the lower-degree candidates, especially when those high-degree candidates are related to the querying user. We also find that if the querying users provide more social connections, the search performance would increase steadily.

We do acknowledge that there are still several limitations in this paper. First of all, our simulation study assumed that the purpose of each query is to find the best-matching candidates so we didn't differentiate the deeper intentions of different queries. However, it is observed that users may develop different strategies in their search processes so that some queries may be only used to filter out certain non-relevant ones. Identifying the search intentions behind each query would give us better understanding of the impacts of privacy concerns. This is one future direction.

Secondly, we also assumed that each querying user is able to tune the optimized configurations of the weights for each feature; while it may not be the case in a live search system. Users may exhibit different behaviors as we expected -- they may not necessary to tune for the optimal parameters and find the best matched candidates. Our next step is to conduct a live user experiment to study how users interact with the search system under different privacy-preserving networks.

Finally, we tested the impacts of local and global network features separately; whereas we know that privacy concerns affect people search system such as in PeopleExplorer [2] on both features. In addition, we studied the privacy settings for the querying users and candidates separately. In the real settings, all these factors should be studied together.

---

[2] http://crystal.exp.sis.pitt.edu:8080/PeopleExplorer/